# Altering control modes of complex networks based on edge removal

Xizhe Zhang[1,2*], Qian Li[1]

[1] (School of Computer Science and Engineering, Northeastern University, Shenyang, Liaoning, China)

[2] (Joint Laboratory of Artificial Intelligence and Precision Medicine of China Medical University and Northeastern University, Shenyang, Liaoning, China)

**Abstract:** Controlling a complex network is of great importance in many applications. The network can be controlled by inputting external control signals through some selected nodes, which are called input nodes. Previous works found that the majority of the nodes in dense networks are either the input nodes or not, which leads to the bimodality in controlling the complex networks. Due to the physical or economic constraints of many real control scenarios, altering the control mode of a network may be critical to many applications. Here we develop a graph-based algorithm to alter the control mode of a network. The main idea is to change the control connectivity of nodes by removing carefully selected edges. We rigorously prove the correctness of our algorithm and evaluate its performance on both synthetic and real networks. The experimental results show that the control mode of a network can be easily changed by removing few selected edges. Our methods provide the ability to design the desired control mode for different control scenarios, which may be useful in many applications.

**Keyword:** complex network; structural controllability; control modes; edge removal; input nodes

## 1. Introduction

The control of the complex networked systems is of great importance in many applications. Recent developments about structural controllability of complex networks[1-3] have attracted much attention. Considerable research efforts have been devoted to analyzing control principle of various real-world networks. For example, network controllability has been used to analyzing the interbank loan network[4,5], finding the underlying mechanism of human brain network[6-8], identifying disease genes[9-13] and finding drug target[13-15].

A network is said to be controllable if its state can be controlled from an initial state to a desired final state in finite time[2]. The nodes used to input the external control signals are called actuator nodes, and the actuator nodes that do not share input signals are called driver nodes[16-21]. The minimum set of driver nodes is called **M**inimum **D**river nodes **S**et (*MDS*). Previous works[22] found that the *MDS* can be obtained by the maximum matching (the maximum edges set which do not share nodes) of the network. Based on this framework, Liu et.al[18] applied the structural controllability theory to the complex networks and found that the size of the *MDS* is closely related to the degree distribution of the networks. Ruths et.al[23,24] further investigated the formation of the *MDS* and found that *MDS* is mainly composed of the source nodes and the sink nodes of the control path. Menichetti et.al[25] found the size of the *MDS* is mainly determined by the number of the low in-degree nodes, especially those with one or two in-degree.

For most of the real networks, the maximum matching is usually not unique because of their structural complexity. Therefore, there may exist many *MDS*s which can fully control the network[16,19]. According to the participation of the nodes in all *MDS*s, a node is called an input node if it appears in at least one *MDS*. Otherwise, it is called redundant node. By using above control type of nodes, many works have been done to investigate the types of nodes in control. For example, input nodes can be used for identifying critical regulatory genes[26], finding cancer-associated genes[20], identifying novel disease genes and potential drug targets[9,10,13].

Previous work[19] found an interesting bimodality phenomenon in dense networks, which is that the majority of the nodes must be either input nodes or redundant nodes. This leads to two control modes of complex networks, centralized and distributed control[19]. For the networks with centralized control, most of the nodes are redundant nodes, which means the selection of driver nodes are very limited. For the networks with distributed control, most of the nodes are input nodes, which suggests that there are many available options to input the external control signals into the network. Different control modes may have different practical applications, for example, distributed control mode may improve the resilience to control the network yet induce potential security risk[19]. Although our previous work[16] provides a method to alter the control modes by adding edges, we still do not know how to alter the control mode by removing edges. Therefore, it is of great practical significance to investigate how to design or alter the control mode of a network.

Here we present an efficient method to alter the control mode of a network by removing edges. The method we presented here is motived by our previous findings that the input nodes and redundant nodes are connected by the alternating paths, respectively[16]. We prove that by altering the control type of few selected nodes, the control mode of a network can be easily changed. The experimental results on synthetic and real networks show the efficiency of our algorithm. Comparing other strategies such as adding edges[17] or reversing the direction of edges[19], the strategy of removing edges may be more feasible in many applications. For example, when controlling the biological network such as protein-protein interaction networks[9], it is impractical or difficult to add new interaction between proteins. However, removing edges can be easily done by inhibiting the signals transmitted between proteins.

## 2. Structural controllability and maximum matching

Consider a linear time-invariant network $G(V, E)$, its dynamics can be described by the following equation:

$$\dot{x}(t) = Ax(t) + Bu(t) \tag{1}$$

where the state vector $x(t)=(x_1(t), …, x_N(t))^T$ denotes the value of $N$ nodes in the network at time $t$, $A$ is the transpose of the adjacency matrix of the network, $B$ is the input matrix that defines how control signals are input to the network, and $u(t)=(u_1(t), …, u_H(t))^T$ represents the $H$ input signals at time $t$. To analyze the controllability of the directed network, we need to convert the directed network to an undirected bipartite graph. The bipartite graph is built by splitting the node set $V$ into two node sets $V^-$ and $V^+$, where a node $n$ in $G$ is converted to two nodes $n^-$ and $n^+$ in $B$. The nodes $n^-$ and $n^+$ are, respectively, connected to the in-edges and out-edges of node $n$.

Next, we introduce some concepts about maximum matching. A set of edges is called a matching if no two edges in the matching share common nodes. The edges belong to a matching are called matched edges. A node is said to be matched if there is a matched edge linked to the node; Otherwise, the node is unmatched. The unmatched nodes in $V^+$ are called unsaturated nodes, and the unmatched nodes in $V^-$ are called driver nodes. A path is said to be an alternating path if the edges of the path are alternately in and not in the matching. An alternating path that begins and ends on the unmatched nodes is called augmenting path. A maximum matching is a matching with the maximum number of edges. According to the structural controllability theory, for any maximum matching of the network, the set of unmatched nodes in $V^-$ is called Minimum Driver nodes Set (*MDS*), which can be used to control the network.

Because maximum matching is not unique for most networks, there may exist numerous MDSs in a network. A node is called an input node if it appears in at least one *MDS*. Otherwise, we call it as a redundant node. Figure 1 shows a simple network with its maximum matching and the node classification. Input nodes and redundant nodes are not evenly distributed in the networks. Previous works[19] found that in some dense networks, the majority of nodes of a network are either input nodes or redundant nodes. This bifurcation phenomenon leads to two control modes of the networks: centralized control and distributed control. For networks with distributed control, most of the nodes are input nodes. For the networks with centralized control, most of the nodes are redundant nodes. Therefore, to alter the control modes of a network, we need to change the control type of most of the nodes of a network.

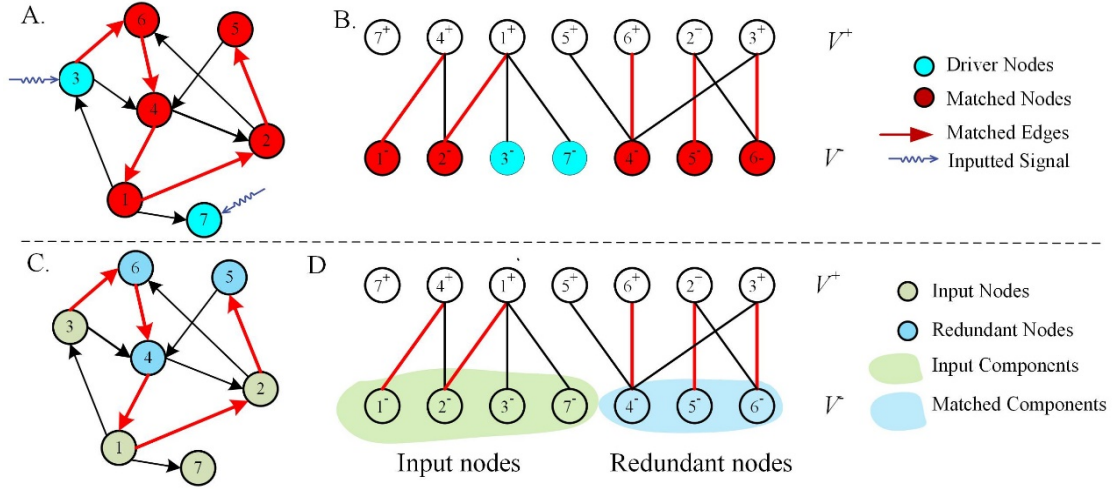

**Figure.1** | A simple network and its node classification. (A). a simple network and one of its maximum matching (red edges). We show its driver nodes (shaded in blue) and matched nodes (shaded in red); (B). The bipartite graph of the network. Each node of the original directed network are split into two nodes, which connected to in and out edges, respectively; (C). Input nodes and redundant nodes of the network. The blue nodes are redundant nodes, and the green nodes are input nodes. The input nodes {1, 2} can be reached by driver nodes {3, 7} through alternating paths. The redundant nodes {4, 5, 6} cannot be reached by any input nodes.

## 3. From distributed control to centralized control

First, we introduce how to change a network from distributed control to centralized control. This is equivalent to altering the control type of most of the nodes from input nodes to redundant nodes. Our previous work[16] found that a node is an input node if and only if it can be reached by an alternating path from a driver node; otherwise, it is a redundant node. Figure 1C give an example of input nodes and their connectivity. For example, the input nodes 1 and 2 can be reached by driver nodes 3 and 7 through alternating paths. However, the redundant nodes 4, 5 and 6 cannot be reached by any driver node through any alternating path. The detailed theorem is as follows:

**Property** 1: For any *MDS D* and a driver node $n \in D$, all nodes of $C(n)$ must be input nodes, where $C(n)$ is the set of all nodes which can be reached from node $n$ through any alternating path.

**Property** 2: For any *MDS D*, if node $m$ cannot be reached by any alternating path start from any driver node of $D$, then $m$ must be a redundant node and never appears in any *MDS*.

Because the input nodes are connected to driver nodes by alternating paths, we can define connected components based on the alternating path, which are called control components. Based on above properties, we now present a new theorem for altering a connected component which only contains input nodes to the component only contains redundant nodes:

**Theorem 1**: For a network $G(V, E)$ and one of its *MDS D*, consider a control component $C$, which contains at least one driver node, all nodes of the set $C' = C - (C \cap D)$ are redundant nodes of network

$G'(V, E')$, where $E' = E - e_{in}(C \cap D)$.

**Proof:** Based on property 1, the input nodes are connected to at least one driver nodes by alternating paths. Therefore, if we disconnect all the driver nodes from the alternating paths, the input nodes will be turned into redundant nodes. The easiest way is to remove all in-edges $e_{in}(C \cap D)$ of the driver nodes. Because driver node are unmatched nodes, therefore, all edges of $e_{in}(C \cap D)$ are unmatched. Therefore, removing the edges $e_{in}(C \cap D)$ will not change the maximum matching. For new network $G'(V, E')$, any driver node of $D$ cannot reached the nodes set $C' = C - (C \cap D)$ by any alternating paths. Based on property 2, all node of set $C' = C - (C \cap D)$ are redundant nodes. The proof is complete.

Figure 2 gives an example subgraph of the whole network for altering input nodes to redundant nodes. By removing the in-edge e(2,5) of driver node 5, the input nodes 3 and 4 of the original network are changed into redundant nodes in the new network. Because we only remove the unmatched edges from the network, the *MDS D* of original network $G(V, E)$ is also an *MDS* of $G'(V, E')$. It means that the control schemes of the network will not change even after changing the control mode of the network. It is one of the advantages of our algorithm.

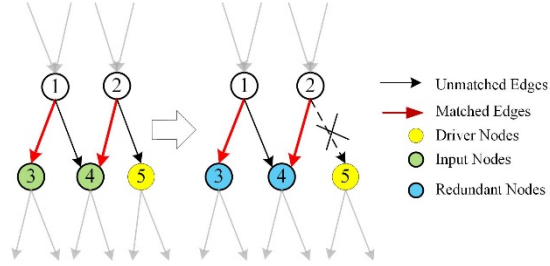

**Figure 2** | Illustration of altering input nodes to redundant nodes. By removing the in-edge of driver node 5, the input nodes 3 and 4 are changed into redundant nodes.

Based on the theorem 1, we designed an efficient algorithm to alter the control mode of a network. The basic idea of the algorithm is to first find the largest connected component of input nodes based on alternating paths and disconnected the driver nodes from the connected component. The above idea and steps are formulated in Algorithm 1 for altering a network from distributed control to centralized control.

---
**Algorithm 1: Altering a network from distributed control to centralized control**

1. **Input**: Network G(*V*, *E*);
2. **Output**: removed edges set $R_e$
3. Find a maximum matching *M* of *G*; let the nodes set without matched in-edges be *D*;
4. **for** each *i*∈*D* **do**
5.     find all alternating paths start from node *i*;
6.     let $P_i$ be the set of nodes which have even distance to node *i* on these alternating paths;
7. **end**
8. **for** each pair *i*≠*j* **do**
9.     **if** $P_i \cap P_j \neq \emptyset$ **then** $P_i = P_j \cup P_i$, $P_j = \emptyset$;
10. **end**
11. Let the largest nodes set be $P_{max}$ and $D_p = P_{max} \cap D$;
12. **for** each node $n_i$ in $D_p$ **do**
13.     $R_e = R_e \cup e_{in}(n_i)$;
14. **end**
15. **Output**: removed edges set $R_e$.
---

## 4. From centralized control to distributed control

Next, we show how to change a network from centralized control to distributed control. This is equivalent to alter the control type of most nodes from redundant nodes to input nodes. Based on property 2, the redundant nodes cannot be reached by any alternating path start from any driver nodes. Therefore, to alter the control type of redundant nodes, we need to change some matched nodes to driver nodes. That can be done by removing carefully selected edges from the alternating paths which start from the redundant nodes.

The edges of an alternating path are alternately in and not in a maximum matching. If the matching is maximum, the alternating paths must be connected to at most one unmatched node. Therefore, there are three possible cases for the alternating paths: case 1 (Figure 3A): alternating path with odd length, where the path starts with a matched node and end with an unmatched node; case 2 (Figure 3B): alternating cycles, where the alternating path form a cycle; case 3 (Figure 3C): alternating path with odd length, where the start and end nodes are both matched;

Figure 3 shows the results when removing a matched edge from an alternating path. For case 1, if we remove a matched edge, the alternating path will be even length. For case 2, if we remove a matched edge, the alternating cycle will turn into alternating paths with even length. However, the type of nodes within the alternating cycle remains unchanged. For case 3, if we remove a matched edge, there will exist a driver node in the alternating path, which makes all nodes on this path be the input nodes based on property 1.

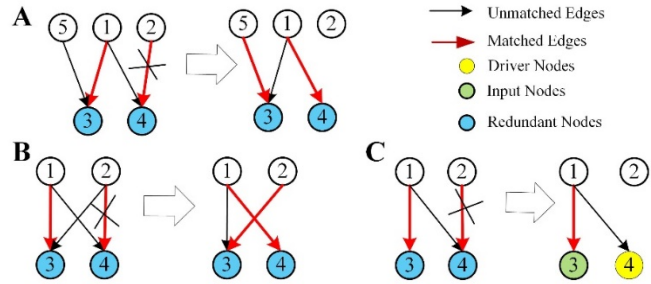

**Figure 3** | Three cases when removing a matched edge from a network. (A). Case 1: alternating paths with an odd length. If we remove a matched edge, the alternating path will be even length; (B). Case2: alternating cycles. If we remove a matched edge, the alternating cycle will turn into alternating paths with even length; (C). Case 3: alternating paths with even length. If we remove a matched edge, there will exist a driver node in the alternating path, which makes all nodes on this path be the input nodes;

Therefore, to change a network from centralized control to distributed control, we need to first find the largest connected components of redundant nodes and change the control type of the nodes. The basic idea is first to break the cycles and remove the unmatched nodes in alternating paths, and change a matched node to driver node by removing its matched edge. Therefore, the other redundant nodes which connected by the driver node will be changed to input nodes. The above idea and steps are formulated in Algorithm 2 for altering a network from centralized control to distributed control.

| | **Algorithm 2: Altering a network from centralized control to distributed control** |
|---|---|
| 1. | **Input**: Network $G(V, E)$; |
| 2. | **Output**: removed edges set $R_e$ |
| 3. | Find a maximum matching $M$ of $G$; let the nodes set without matched out-edges be $U$; |
| 4. | **for** each $i \in U$ **do** |
| 5. |    find all alternating paths start from node $i$; |
| 6. |    let $P_i$ be the set of nodes which have even distance to node $i$ on these alternating paths; |
| 7. | **end** |
| | //detach all unmatched nodes from redundant nodes |
| 8. | **for** each pair $i \neq j$ **do** |

9.        **if** $P_i \cap P_j \neq \emptyset$, let $P_i = P_j \cup P_i$, $P_j = \emptyset$;
10.    **end**
11.    Find the largest nodes set $P_{max}$; Let $U_p = P_{max} \cap U$;
12.    **for** each node $n_i$ in $U_p$ do
13.        $R_e = R_e \cup e_{out}(n_i)$;
14.    **end**
     //remove cycles
15.    find all alternating cycles $C_i$ of $P_{max}$;
16.    **for** each $c_j \in C_i$ do
17.        select one matched edge $e_j \in c_j$; $R_e = R_e \cup e_j$;
18.    **end**
     // remove one matched edge
19.    **for** each node $n$ in $P_{max}$ do
20.        find all alternating paths start from node $n$;
21.        let $H_i$ be the set of nodes which have odd distance to node $n$ on these alternating paths;
22.    **end**
23.    Find the largest nodes set $H_k$, let its matched edge is $e_k$, let $R_e = R_e \cup e_k$;
24.    **Output**: removed edges set $R_e$.

## 5. Results

To investigate the efficiency of our methods, we constructed a series of networks based on scale-free network model[27]. The power exponent of in-degree distribution and out-degree distribution are $\gamma_{in}=\gamma_{out}=3$. The average degree $<k>$ is from 5 to 40 with increment 0.1. For each $<k>$, we generated 100 random instances. Therefore, there are total 35,000 network instances used to evaluate the methods.

First, we computed the percentage of driver nodes $N_D$ (Figure 4A), the percentage of input nodes $I_D$ (Figure 4B), the size of the maximum connected component of input nodes (Figure 4C) and the size of the maximum connected component of redundant nodes (Figure 4D) for each network. With the increase of the average degree of the networks, we can observe that the percentage of driver nodes monotonically decreased (Figure.4A), indicating the denser networks are easier controlled. However, the percentage of input nodes exhibited a bifurcation phenomenon, indicating that for the dense network, most nodes are either input nodes or redundant nodes (Figure.4B). Furthermore, these nodes are connected by alternating paths and constitute the large connected components (Figure 4C-4D). Therefore, to alter the control modes of the network, we only need to change the type of the maximum connected component of nodes.

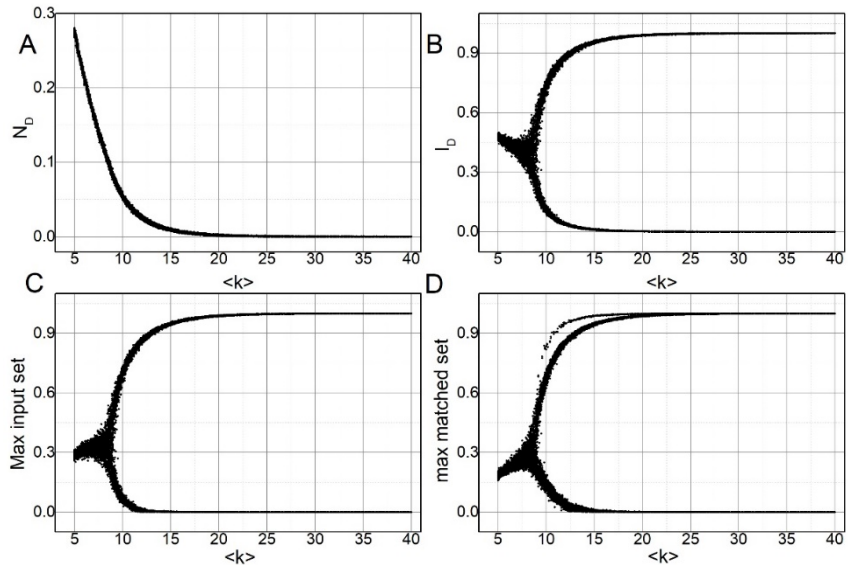

**Figure.4** | Percentage of driver nodes and input nodes of Scale-free networks with $\gamma_{in}=\gamma_{out}=3$. A. percentage of driver nodes $N_D$ versus average degree; B. percentage of input nodes $I_D$ versus average degree; C. percentage of maximum connected component of input nodes versus average degree; D. percentage of maximum connected component of redundant nodes versus average degree.

Next, we evaluated the performance of our algorithms. Figure.5A shows an example for changing a network from distributed mode to centralized mode. By removing only 11 edges, almost all input nodes of the original network are changed to redundant nodes. For more complex networks, our algorithms are still efficient to change the control modes of the networks. For networks with different average degrees, we computed the percentage of removed edges and the changed input nodes after removing these edges. Figure 5B shows that with the increase of $k$, very few removed edges can significant changed the type of most input nodes. For example, for dense networks ($k>11$), only 0.5% removed edges can change the type of nearly 80% nodes. For sparse networks ($k<8$), we can still change almost 30% nodes with less than 8% edges. Note that the number of input nodes of these networks is also small ($I_D<0.4$, Figure 6B), which means that our algorithm can change many input nodes even in sparse networks. Furthermore, we evaluated the efficiency of our algorithm by computing the percentage of changed input nodes per removed edge $\Delta n_D/p$. Figure 5C shows that $\Delta n_D/p$ increases rapidly with the average degree $k$, which indicates that it is easier to change the control mode for dense networks. Above results are rooted that in the dense networks, most nodes are connected by the alternating paths and form a large connected component (Figure 4C-4D), that means to alter the unmatched nodes of a connected component will also alter the other nodes of the connected set. Therefore, the larger the connected component is, the easier networks are changed. For the networks with centralized modes, the results are similar (Figure 6). Although the algorithm needs more steps to change from centralized control to distributed control, the resulted are not significantly increased. Therefore, we can also easily change a network from centralized modes to distributed modes.

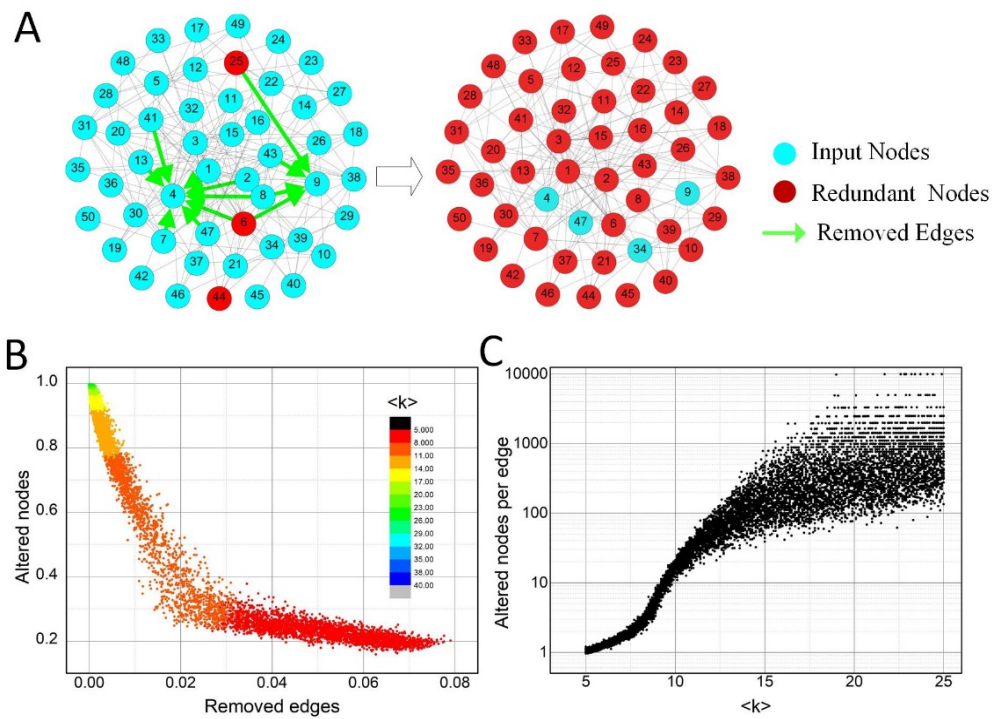

**Figure 5** | Changing a network from distributed control to centralized control. **A**. an example network with distributed control. By removing 11 edges (green edges), most nodes are changed to redundant nodes. **B**. percentage of removed edges versus percentage of the altered input node. **C**. number of altered input nodes per edges. The networks we used are generated by scale-free networks model with degree exponents $\gamma_{in}=\gamma_{out}=3$, $N=10^4$ and average degree $<k>\in[5,25]$. For each $<k>$, we generate 100 networks.

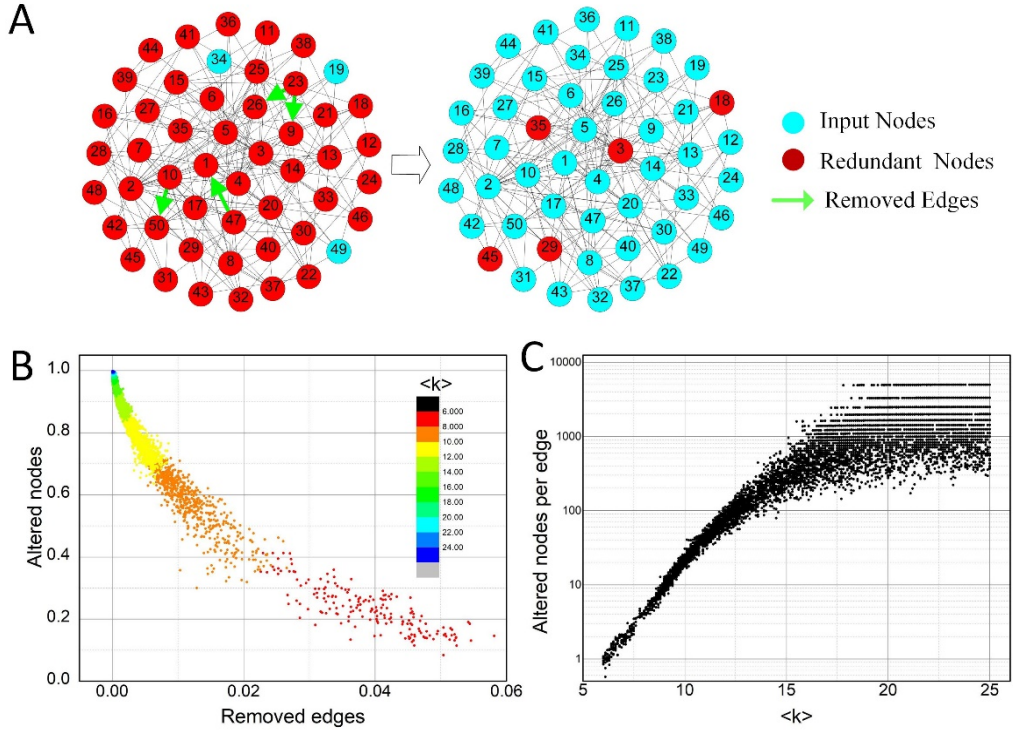

**Figure 6** | Changing a network from centralized control to distributed control. **A**. an example network with distributed control. By removing 4 edges (green edges), most nodes are changed to redundant nodes. **B**. percentage of removed edges versus percentage of the altered input node. **C**. number of altered input nodes per edges. The networks we used are generated by scale-free networks model with degree exponents $\gamma_{in}=\gamma_{out}=3$, $N=10^4$ and average degree $<k>\in[5,25]$. For each $<k>$, we generate 100 networks.

Next, we evaluated the performance of our algorithms on real networks. These networks are selected based on their diversity of topological structure, including biological networks, social networks, and technical networks. For each network, we show its type, name, number of nodes (*N*) and edges (*L*), density of driver nodes $n_{MDS}$, size and type of the largest connected component $CC_{max}$, in which *I*, *R* denote input components and redundant components respectively, the proportion of edges (*p*) that is removed from the network to change the type of $CC_{max}$ and the density of changed input nodes ($\Delta n_D$) after removing edges. The detailed results are listed in Table.1. For networks with the large connected components, we can change their control modes by removing very few edges. For example, for Amazon co-purchasing network, only 2% removed edges can change 80% or more input nodes of the network. However, for some networks with the small connected components, such as s208a, s420a, s838a, it is hard to change their control types. The reason may root that the size of connected components of these networks is on average small; therefore altering the maximum connected components is not enough to change the control mode of the networks. In this case, we need to remove more edges to change the control mode of the networks.

**Table 1**. Results for real networks. For each network, we show its type, name, number of nodes ($N$) and edges ($L$), density of driver nodes $n_{MDS}$, size and type of the largest control component $CC_{max}$, in which $I$, $M$ denote Input set and Matched set respectively, the proportion of edges ($p$) that is removed from the network to change the type of $CC_{max}$ and the density of changed driver nodes ($\Delta n_D$) after removing edges.

| Type | Name | $N$ | $L$ | $<k>$ | $n_{MDS}$ | $CC_{max}$ | $p$ | $\Delta n_D$ |
|---|---|---|---|---|---|---|---|---|
| **Food Web** | StMarks | 54 | 356 | 13.19 | 24.07% | 38.89%(R) | 10.39% | 62.96% |
| | Mangrove | 97 | 1492 | 30.76 | 22.68% | 55.67%(I) | 5.63% | 34.02% |
| | Silwood | 154 | 370 | 4.81 | 75.32% | 84.42%(I) | 75.14% | 14.29% |
| **Trust** | Prison inmate | 67 | 182 | 5.43 | 13.43% | 59.70%(R) | 6.59% | 64.18% |
| | Slashdot0902 | 82168 | 948464 | 23.09 | 4.55% | 91.23%(I) | 9.27% | 86.68% |
| | WikiVote | 7115 | 103689 | 29.15 | 66.56% | 32.12%(R) | 22.69% | 25.76% |
| **Electronic circuits** | s208a | 122 | 189 | 3.10 | 23.77% | 17.21%(I) | 6.35% | 4.10% |
| | s420a | 252 | 399 | 3.17 | 23.41% | 9.13%(I) | 3.51% | 1.98% |
| | s838a | 512 | 819 | 3.20 | 23.24% | 5.27%(I) | 2.20% | 0.98% |
| **Citation** | ArXiv-HepTh | 27770 | 352807 | 25.41 | 21.58% | 48.96%(R) | 4.56% | 57.61% |
| | SciMet | 3084 | 10416 | 6.75 | 37.48% | 52.14%(I) | 20.32% | 29.83% |
| | Google | 875713 | 5105039 | 11.66 | 36.95% | 60.80%(I) | 4.20% | 10.37% |
| **Internet** | p2p-1 | 10876 | 39994 | 7.35 | 55.20% | 90.58%(I) | 49.92% | 35.85% |
| | p2p-2 | 8846 | 31839 | 7.20 | 57.78% | 90.55%(I) | 50.82% | 34.69% |
| | p2p-3 | 8717 | 31525 | 7.23 | 57.74% | 91.75%(I) | 50.85% | 35.41% |
| **Organizational** | Consulting | 46 | 879 | 38.22 | 4.35% | 97.83%(I) | 3.53% | 93.48% |
| **Social communication** | UClonline | 1899 | 20296 | 21.38 | 32.33% | 79.94%(I) | 11.87% | 49.55% |
| **Product co-purchasing** | Amazon0302 | 262111 | 1234877 | 9.42 | 3.23% | 49.55%(R) | 1.41% | 74.78% |
| | Amazon0312 | 400727 | 3200440 | 15.97 | 3.52% | 83.61%(R) | 0.21% | 81.52% |
| | Amazon0505 | 410236 | 3356824 | 16.37 | 3.62% | 91.35%(I) | 0.92% | 87.78% |
| | Amazon0601 | 403394 | 3387388 | 16.79 | 2.04% | 75.90%(R) | 1.33% | 90.65% |
| **Social network** | twitter_combined | 81306 | 1768149 | 43.49 | 19.39% | 79.40%(I) | 3.03% | 60.36% |
| | Facebook_0 | 347 | 5038 | 29.04 | 5.48% | 86.46%(R) | 0.04% | 81.84% |
| | Facebook_107 | 1912 | 53498 | 55.96 | 45.92% | 54.08%(R) | 0.00% | 53.24% |
| | Facebook_348 | 572 | 6384 | 22.32 | 61.01% | 38.64%(R) | 0.03% | 38.29% |

Finally, we apply our algorithms to human **P**rotein-**P**rotein **I**nteraction (**PPI**) network. Previous works[28] have found that some redundant nodes of PPI network are related to cancer genes and drug targets. Altering the control type of these nodes may play an important role in the transition from healthy to disease states. We used a directed human PPI network[29] which contains 6,339 proteins and 34,813 interactions. The PPI network has 3707 input nodes and 2631 redundant nodes, therefore, the control mode of the network are distributed control. We used algorithm 1 to change the network from distributed control to centralized control. After removing 11.8% (4108 of 34813) edges, the redundant nodes are increased from 2631 to 4011, which means the control mode of the new network is centralized control.

The detailed list of removed edges can be found in the supplemental material. Furthermore, by comparing online databases such as Online Mendelian Inheritance in Man (OMIM) (omim.org), we found that most of the removed edges are related to disease genes. For example, 48.3% (1986 of 4108) removed edges have two nodes related to disease genes, and 40.8% (1675 of 4108) edges have one node related to disease genes. This may provide us a new way to understand the underlying principle of healthy state and disease state of PPI network.

## 6. Conclusion

Controlling complex networks is of great importance in many applications. How to change the control modes of complex networks is important in many real control scenarios. There may exist several ways to modify the topological structure of complex networks, such as add or remove nodes and edges, reverse the direction of edges. In many real networks, removing edges may be feasible because it can be done by simply blocking the signals in the edges.

Here we present a novel algorithm to alter the control mode of a network based on edge removal. The main idea is to alter the connectivity of nodes in control based on edge removal. One advantage of our algorithm is that when we alter a network from distributed control to centralized control, the control schemes of the network will not change. We evaluated the performance of our algorithms in both synthetic and real networks, the results showed that the control mode of the denser network can be easily changed by removing very few carefully selected edges. Furthermore, for human Protein-Protein Interaction (**PPI**) network, we found that most of the removed edges are related to the disease genes, which provide us a new way to understand the underlying principle between healthy state and disease state.

These findings will improve our understanding of the control principles of complex networks and may be useful in various real control scenarios.

**Acknowledgments**

Supported by the Natural Science Foundation of China under grant number 60093009, 91546110.

**Author Contributions**

X.-Z.Z. designed research wrote the paper and analyzed the data. Q.L. performed the experiments. All authors reviewed the manuscript.

**Additional Information**

The authors declare no competing financial interests. Correspondence and requests for materials should be addressed to X.-Z.Z. (Email: zhangxizhe@mail.neu.edu.cn)

## References

[1] Lin, C.-T. Structural controllability. *IEEE Transactions on Automatic Control* 19, 201-208 (1974).
[2] Kalman, R. E. Mathematical description of linear dynamical systems. *Journal of the Society for Industrial and Applied Mathematics, Series A: Control* 1, 152-192 (1963).
[3] Luenberger, D. Introduction to dynamic systems: theory, models, and applications.(1979).
[4] Delpini, D. *et al.* Evolution of controllability in interbank networks. *Scientific reports* 3, 1626 (2013).
[5] Razakanirina, R. M. & Chopard, B. Risk analysis and controllability of credit market. *ESAIM: Proceedings and Surveys* 49, 91-101 (2015).
[6] Kumar, A., Vlachos, I., Aertsen, A. & Boucsein, C. Challenges of understanding brain function by

selective modulation of neuronal subpopulations. *Trends in neurosciences* 36, 579-586 (2013).

[7] Gu, S. *et al.* Controllability of structural brain networks. *Nature communications* 6 (2015).

[8] Tang, E. & Bassett, D. S. Control of Dynamics in Brain Networks. *ArXiv preprint 1701.01531* (2017 ).

[9] Wuchty, S. Controllability in protein interaction networks. *Proceedings of the National Academy of Sciences* 111, 7156-7160 (2014).

[10] Amand, M. M. S., Tran, K., Radhakrishnan, D., Robinson, A. S. & Ogunnaike, B. A. Controllability analysis of protein glycosylation in CHO cells. *PloS one* 9, e87973 (2014).

[11] Zhang, X.-F., Ou-Yang, L., Zhu, Y., Wu, M.-Y. & Dai, D.-Q. Determining minimum set of driver nodes in protein-protein interaction networks. *BMC bioinformatics* 16, 146 (2015).

[12] Ishitsuka, M., Akutsu, T. & Nacher, J. C. Critical controllability in proteome-wide protein interaction network integrating transcriptome. *Scientific reports* 6 (2016).

[13] Vinayagam, A. *et al.* Controllability analysis of the directed human protein interaction network identifies disease genes and drug targets. *Proc Natl Acad Sci U S A* 113, 4976-4981, doi:10.1073/pnas.1603992113 (2016).

[14] Asgari, Y., Salehzadeh-Yazdi, A., Schreiber, F. & Masoudi-Nejad, A. Controllability in cancer metabolic networks according to drug targets as driver nodes. *PLoS One* 8, e79397 (2013).

[15] Wu, L., Shen, Y., Li, M. & Wu, F.-X. Network output controllability-based method for drug target identification. *IEEE transactions on nanobioscience* 14, 184-191 (2015).

[16] Zhang, X., Lv, T. & Pu, Y. Input graph: the hidden geometry in controlling complex networks. *Scientific reports* 6 (2016).

[17] Zhang, X., Wang, H. & Lv, T. Efficient target control of complex networks based on preferential matching. *PloS one* 12, e0175375 (2017).

[18] Liu, Y.-Y., Slotine, J.-J. & Barabási, A.-L. Controllability of complex networks. *Nature* 473, 167-173 (2011).

[19] Jia, T. *et al.* Emergence of bimodality in controlling complex networks. *Nature communications* 4 (2013).

[20] Liu, X. & Pan, L. Identifying driver nodes in the human signaling network using structural controllability analysis. *IEEE/ACM Transactions on Computational Biology and Bioinformatics (TCBB)* 12, 467-472 (2015).

[21] Liu, Y.-Y. & Barabási, A.-L. Control principles of complex systems. *Reviews of Modern Physics* 88, 035006 (2016).

[22] Murota, K. *Matrices and matroids for systems analysis*. Vol. 20 (Springer Science & Business Media, (2009).

[23] Ruths, J. & Ruths, D. Control profiles of complex networks. *Science* 343, 1373-1376 (2014).

[24] Campbell, C., Ruths, J., Ruths, D., Shea, K. & Albert, R. Topological constraints on network control profiles. *Scientific reports* 5 (2015).

[25] Menichetti, G., Dall'Asta, L. & Bianconi, G. Network controllability is determined by the density of low in-degree and out-degree nodes. *Physical review letters* 113, 078701 (2014).

[26] Ravindran, V., Sunitha, V. & Bagler, G. Identification of critical regulatory genes in cancer signaling network using controllability analysis. *Physica A: Statistical Mechanics and its Applications* 474, 134-143 (2017).

[27] Goh, K.-I., Kahng, B. & Kim, D. Universal behavior of load distribution in scale-free networks. *Physical Review Letters* 87, 278701 (2001).

[28] A. Vinayagam, T. E. Gibson, H. J. Lee, B. Yilmazel, C. Roesel, Y. Hu, Y. Kwon, A. Sharma, Y. Y.

Liu, N. Perrimon, and A. L. Barabasi, Controllability analysis of the directed human protein interaction network identifies disease genes and drug targets, *Proc Natl Acad Sci U S A,* vol. 113, no. 18, pp. 4976-81, May 03, 2016.

[29] A. Vinayagam, U. Stelzl, R. Foulle, S. Plassmann, M. Zenkner, J. Timm, H. E. Assmus, M. A. Andradenavarro, and E. E. Wanker, A directed protein interaction network for investigating intracellular signal transduction, *Science Signaling,* vol. 4, no. 189, pp. rs8-rs8, 2011.

[30] C. Knox, V. Law, T. Jewison, P. Liu, S. Ly, A. Frolkis, A. Pon, K. Banco, C. Mak, and V. Neveu, DrugBank 3.0: a comprehensive resource for 'omics' research on drugs, *Nucleic acids research,* vol. 39, no. suppl 1, pp. D1035-D1041, 2011.